\documentclass[
 reprint,
superscriptaddress,
amsmath,amssymb,
prl,
]{revtex4-2}

\usepackage{graphicx}
\usepackage{dcolumn}
\usepackage{bm}
\usepackage{hyperref}
\usepackage{color}
\usepackage[mathlines]{lineno}
\usepackage{tabularx}
\usepackage{subfigure}
\usepackage[normalem]{ulem}
\usepackage[export]{adjustbox}
\usepackage{url}
\usepackage{lineno}
\usepackage[flushleft]{threeparttable}
\usepackage{makecell}
\usepackage{multirow}
\usepackage{rotating}

\hypersetup{
    pdfnewwindow=true,    
    colorlinks=true,      
    linkcolor=blue,       
    citecolor=blue,       
    filecolor=blue,       
    urlcolor=blue         
}

\usepackage{natbib} 

\usepackage{url}
\usepackage{ulem}
\usepackage{amsmath}
\usepackage{amsfonts}
\usepackage{textcomp}
\usepackage{subfigure}
\usepackage{verbatim}
\usepackage{rotating}
\usepackage{tikz}
\usepackage{graphicx}
\usepackage{color}
\usepackage{epstopdf}

\begin{document}

\title{\bf \boldmath  All-Aky Search for Individual Primordial Black Hole Bursts with LHAASO}

\author{Zhen Cao}
\affiliation{Key Laboratory of Particle Astrophysics \& Experimental Physics Division \& Computing Center, Institute of High Energy Physics, Chinese Academy of Sciences, 100049 Beijing, China}
\affiliation{University of Chinese Academy of Sciences, 100049 Beijing, China}
\affiliation{TIANFU Cosmic Ray Research Center, Chengdu, Sichuan,  China}
 
\author{F. Aharonian}
\affiliation{TIANFU Cosmic Ray Research Center, Chengdu, Sichuan,  China}
\affiliation{University of Science and Technology of China, 230026 Hefei, Anhui, China}
\affiliation{Yerevan State University, 1 Alek Manukyan Street, Yerevan 0025, Armenia}
\affiliation{Max-Planck-Institut for Nuclear Physics, P.O. Box 103980, 69029  Heidelberg, Germany}
 
\author{Y.X. Bai}
\affiliation{Key Laboratory of Particle Astrophysics \& Experimental Physics Division \& Computing Center, Institute of High Energy Physics, Chinese Academy of Sciences, 100049 Beijing, China}
\affiliation{TIANFU Cosmic Ray Research Center, Chengdu, Sichuan,  China}
 
\author{Y.W. Bao}
\affiliation{Tsung-Dao Lee Institute \& School of Physics and Astronomy, Shanghai Jiao Tong University, 200240 Shanghai, China}
 
\author{D. Bastieri}
\affiliation{Center for Astrophysics, Guangzhou University, 510006 Guangzhou, Guangdong, China}
 
\author{X.J. Bi}
\affiliation{Key Laboratory of Particle Astrophysics \& Experimental Physics Division \& Computing Center, Institute of High Energy Physics, Chinese Academy of Sciences, 100049 Beijing, China}
\affiliation{University of Chinese Academy of Sciences, 100049 Beijing, China}
\affiliation{TIANFU Cosmic Ray Research Center, Chengdu, Sichuan,  China}
 
\author{Y.J. Bi}
\affiliation{Key Laboratory of Particle Astrophysics \& Experimental Physics Division \& Computing Center, Institute of High Energy Physics, Chinese Academy of Sciences, 100049 Beijing, China}
\affiliation{TIANFU Cosmic Ray Research Center, Chengdu, Sichuan,  China}
 
\author{W. Bian}
\affiliation{Tsung-Dao Lee Institute \& School of Physics and Astronomy, Shanghai Jiao Tong University, 200240 Shanghai, China}
 
\author{A.V. Bukevich}
\affiliation{Institute for Nuclear Research of Russian Academy of Sciences, 117312 Moscow, Russia}
 
\author{C.M. Cai}
\affiliation{School of Physical Science and Technology \&  School of Information Science and Technology, Southwest Jiaotong University, 610031 Chengdu, Sichuan, China}
 
% ...
\collaboration{The LHAASO Collaboration}

\email{jianghb@ihep.ac.cn\\
bixj@ihep.ac.cn\\
hushicong@ihep.ac.cn\\
lizhe@ihep.ac.cn\\
wangsai@ihep.ac.cn\\
zhangxin@mail.neu.edu.cn
}

\author{S. Wang}
\affiliation{Theoretical Physics Division, Institute of High Energy Physics, Chinese Academy of Sciences, Beijing 100049, China}
\affiliation{School of Physics, Hangzhou Normal University, Hangzhou 311121, Zhejiang, China}
\author{Xin Zhang}
\affiliation{Liaoning Key Laboratory of Cosmology and Astrophysics, College of Sciences, Northeastern University, Shenyang 110819, China}
\affiliation{MOE Key Laboratory of Data Analytics and Optimization for Smart Industry, Northeastern University, Shenyang 110819, China}
\affiliation{National Frontiers Science Center for Industrial Intelligence and Systems Optimization, Northeastern University, Shenyang 110819, China}



\begin{abstract}
Primordial black holes~(PBHs) are hypothetical black holes with a wide range of masses that formed in the early Universe. As a result, they may play an important cosmological role and provide a unique probe of the early Universe. 
A PBH with an initial mass of approximately $10^{15}$~g is expected to explode today in a final burst of Hawking radiation. In this work, we conduct an all-sky search for individual PBH burst events using the data collected from March 2021 to July 2024 by the Water Cherenkov Detector Array of the Large High Altitude Air Shower Observatory (LHAASO). Three PBH burst durations, 10~s, 20~s, and 100~s, are searched in the gamma-ray energy range from about 1 TeV to 15 TeV, with no significant PBH bursts observed. The upper limit on the local PBH burst rate density is set to be as low as 181~pc$^{-3}$~yr$^{-1}$ at 99$\%$ confidence level, representing the most stringent limit achieved to date.
\end{abstract}

\oddsidemargin  -0.2cm
\evensidemargin -0.2cm
\maketitle 

$Introduction$$-$Primordial black holes (PBHs) are predicted to form from overdensities during the early Universe through various mechanisms~\cite{Hawking:1971ei,Carr:1974nx,Carr:2020gox}. The PBH mass is related to the cosmological horizon mass at formation, spanning an enormous range from the Planck mass ($10^{-5}$~g) to $10^{5}$ solar mass or larger. PBHs have attracted much attention recently, as they could be responsible for some of the gravitational wave signals from binary black holes detected by the LIGO/Virgo~\cite{Bird:2016dcv,Sasaki:2016jop,Clesse:2016vqa}. 
As PBHs formed in the radiation-dominated era, they are nonbaryonic and could serve as a good dark matter candidate. Over the past decades, the abundance of PBHs has been tightly constrained by numerous cosmological and astrophysical observations.
Nevertheless, even if PBHs constitute only a small fraction
of the dark matter, PBHs could still have important consequences and resolve various cosmological conundra~\cite{Carr:2020xqk,Carr:2021bzv,Escriva:2022duf,Carr:2023tpt}.
In addition, large PBHs could also provide seeds for the supermassive black holes thought to reside in galactic nuclei. 

In 1974, Hawking proposed that black holes can emit all particle species with a black-body spectrum through their lifetime~\cite{Hawking:1974rv}. In the absence of new physics beyond the standard model (SM) of particle physics, a PBH with an initial mass $M$ has a lifetime of $10^{10}(\frac{M}{10^{15}\rm{g}})^3$~yr.
An intriguing aspect of PBHs is that they can be small enough for Hawking radiation to play a significant role. PBHs with initial mass smaller than $10^{15}$~g have already evaporated by now with many important cosmological and astrophysical effects~\cite{Carr:2009jm}.
PBHs with an initial mass around $10^{15}$~g would have a lifetime close to the age of the Universe and would be evaporating at the present epoch. 
As the PBH emits Hawking radiation, its mass decreases while its temperature increases. The final stage of evaporation is expected to produce a burst of high-energy particles, lasting from seconds to tens of seconds, but with an extremely high intensity~\cite{MacGibbon:1991tj,MacGibbon:1990zk,Ukwatta:2015iba,MacGibbon:2007yq}.  The final evaporation provides a unique opportunity to directly detect PBHs at Earth as sudden bursts of gamma rays in the sky.
As an important probe of the early Universe, the observational evidence of a PBH final evaporation event could give significant information about the early Universe, high energy particle physics beyond the energy scale of the Large Hadron Collider~\cite{Evans:2008zzb}, and the convolution of gravitation with thermodynamics. In particular, the final burst should also produce new particles predicted in several extensions of the SM of particle physics. The existence of new particles could significantly modify the final evaporation, making its detection a unique window to explore physics beyond the SM~\cite{Baker:2021btk,Baker:2022rkn,Calza:2023iqa}. Even the nonexistence of PBHs also yields important constraints about the early Universe~\cite{Carr:2009jm}. For instance, limits on the local burst rate density of PBHs can provide tighter limits on the power spectrum of primordial density perturbations at much smaller scales than those probed by the cosmic microwave background~\cite{Wang:2019kaf}, thereby allowing the exploration of various inflation models.


Since PBHs were proposed in the 1970s~\cite{Zeldovich:1967lct,Hawking:1971ei}, no PBH candidates have been observed. Their abundance has been constrained across a variety of mass ranges through various phenomenological effects on the evolution of the Universe and the formation of structures ~\cite{Carr:2020gox,Villanueva-Domingo:2021spv}. In the very-high-energy (VHE) gamma-ray regime, gamma-ray bursts (GRBs) resulting from the final evaporation of individual PBHs have been extensively searched by several GeV/TeV gamma-ray observatories~\cite{Linton:2006yu,Alexandreas:1993zx,Glicenstein:2013vha,Archambault:2017asc,Fermi-LAT:2018pfs,Abdo:2014apa, HAWC:2019wla,HESS:2023zzd}. In searches for final PBH burst timescales from a second to several seconds, the HAWC Observatory established the strongest limit on the local PBH burst rate density at the level of $<3.4\times 10^3$~$\rm{pc}^{-3}\rm{yr}^{-1}$ for PBHs within a distance of 0.5~pc. \textit{Fermi}-LAT conducted a search for PBH evaporation with a remaining lifetime of months to years by detecting the proper motion of gamma-ray point sources~\cite{Fermi-LAT:2018pfs}, thereby deriving a limit on the local PBH evaporation rate in the vicinity
of Earth, at the level of $<7.2\times 10^3$~$\rm{pc}^{-3}\rm{yr}^{-1}$. 
Thanks to its excellent sensitivity and large field of view, the Large High Altitude Air Shower Observatory (LHAASO) offers a significant advantage in the all-sky search for GeV-TeV gamma-ray bursts and is expected to significantly improve constraints on the local PBH burst rate~\cite{Yang:2024vij}.

In this Letter, we conduct an all-sky search for final PBH bursts of VHE gamma-ray emission with timescales of 10~s, 20~s, and 100~s,  using 1241 days of observational data collected by the Water Cherenkov Detector Array (WCDA) of LHAASO. These timescales are chosen as the most sensitive durations to capture the final moments of PBH bursts, with steady gamma-ray sources safely ignored during such a short duration. A detailed sensitivity study of LHAASO for PBH detection is provided in the Supplemental Material~\cite{supp-met}.
Any burst detected in the data is further validated as a PBH candidate by comparing the observed spectrum and light curves with the expected theoretical ones for PBHs to verify consistency.

$LHAASO$ $experiment$$-$The LHAASO experiment
~\cite{Ma:2022aau} is a multipurpose and comprehensive extensive air shower array, located at Mt. Haizi at an altitude of 4410 meters above sea level in Daocheng, Sichuan province, China. It is designed for the study of cosmic rays and gamma rays with energies from sub-TeV to beyond 1~PeV. LHAASO is composed of three subarrays: the KiloMeter Squared Array (KM2A) for gamma-ray astronomy above 10~TeV and cosmic ray physics; WCDA for gamma-rays above a few hundred GeV, and the Wide Field-of-View air Cherenkov Telescopes Array for cosmic ray physics from 10 TeV to 1 EeV. 
The primary goal of WCDA is to survey the northern gamma-ray sky in the VHE band~\cite{LHAASO:2021ozi}. More details about the LHAASO detector are described in
Refs.~\cite{LHAASO:2021ozi,Aharonian:2020iou}. 
Simulations suggest that the sensitivity of WCDA for detecting PBH bursts is more than 10 times higher than that of KM2A. Hence, only the data collected by WCDA are used in this study (details about the simulation are described below).

$Data$ $analysis$$-$Based on the standard evaporation model (SEM)~\cite{MacGibbon:1991tj,MacGibbon:1990zk},  a Monte Carlo (MC) sample of the final PBH bursts is simulated to study the expected signals and to optimize our search strategy. In the SEM, the final burst of a BH is an astronomical burst of photons, neutrinos, electrons, positrons, protons, and antiprotons. The expected gamma-ray spectrum is a superposition of a primary component (directly from Hawking radiation) and a secondary component (from the decay of gauge bosons or heavy leptons, and the hadrons produced by the fragmentation of primary quarks and gluons)~\cite{Ukwatta:2015iba}.  
The time-integrated number of photons per unit energy, $dN_{\gamma}/dE_{\gamma}$, for various burst durations is given by~\cite{Petkov:2008rz}:
\begin{align}
&\frac{d N_\gamma}{d E_\gamma} \approx  \nonumber \\
&9 \times 10^{35}\left\{\begin{array}{cl}
\left(\frac{1~\mathrm{GeV}}{T_{\rm{BH}}}\right)^{3 / 2}\left(\frac{1~\mathrm{GeV}}{E_\gamma}\right)^{3 / 2} \mathrm{GeV}^{-1} & \text { for } E_\gamma<T_{\rm{BH}} \\
\left(\frac{1 \mathrm{GeV}}{E_\gamma}\right)^3 \mathrm{GeV}^{-1} & \text { for } E_\gamma \geq T_{\rm{BH}}
\end{array}\right. \text {,}
\label{eq-2}
\end{align}
where $T_{\rm{BH}}\simeq 7.8 \times 10^{3} \left(\frac{1 s}{\tau}\right) ^{1/3} ~\rm{GeV}$ is the PBH temperature in terms of its
remaining lifetime ($\tau$). 
The expected number of gamma rays ($N_{exp}$) detected by the WCDA as a function of the PBH distance $r$, $\tau$, and zenith angle $\theta$ is given by:
\begin{equation}
    N_{exp}(r, \theta, \tau)=\frac{1}{4 \pi r^2} \int_{E_1}^{E_2} \frac{d N_{\gamma}(\tau)}{d E_{\gamma}} A(E_{\gamma}, \theta) d E_{\gamma},
\end{equation}
where $E_1$ and $E_2$ are the WCDA's energy range, and $A(E_{\gamma},\theta)$ denotes the effective area of WCDA as a function of photon energy and zenith angle. $A(E_{\gamma},\theta)$ is simulated with the G4WCDA, a {\sc geant4}-based~\cite{GEANT4:2002zbu} package developed to simulate the WCDA response.
Assuming a uniform distribution of PBHs in the solar neighborhood, we simulate a random set of PBH burst events under the assumption of a local burst rate density $\rho=5000$~$\rm{pc}^{-3}\rm{yr}^{-1}$, with distances up to $1~\rm{pc}$. No observable signals are expected beyond $1~\rm{pc}$, even for the largest burst duration of $100~\rm{s}$. In the WCDA energy range, the burst emission time profile of PBHs is calculated to be $dN_{\gamma}/{dt} \sim \tau^{-0.53}$ (according to the method described in Sec. 3.5 of Ref.~\cite{Ukwatta:2015iba}), which is used to simulate the light curve of each PBH burst. As the remaining lifetime decreases, the emission rate increases sharply during the final few seconds, resulting in a VHE gamma-ray burst. The spatial distribution of each burst event is simulated by convolving the expected point signal with the angular resolution of the detector, also known as the point spread function (PSF).

Gamma rays and cosmic rays entering the atmosphere produce extensive air showers, which subsequently fire the photomultiplier tubes (PMTs) and are recorded by the WCDA.
To ensure high-quality reconstruction data, each event is required to have at least 30 triggered PMTs and a zenith angle of less than $50^{\circ}$. 
For each shower event, the root mean square distance (RMSD) for shower core fitting is calculated based on the 10 hottest detectors. To achieve better resolution of the shower core position, the RMSD is required to be $\leq 20$~m. The selected events are subsequently categorized into seven groups based on the effective number of triggered PMT units, denoted as $N_{\rm{hit}}$, with ranges: [30 60), [60 100), [100 200), [200 300), [300 500), [500 800), and $\geq$~800. The mean energies corresponding to the $N_{\rm{hit}}$ groups are approximately 1.0, 1.5, 2.1, 2.7, 3.8, 7.1, and 15.0 TeV, simulated using the energy spectrum of $\tau = 20$ s.
To suppress the cosmic-ray background, a gamma-proton discrimination parameter, $\mathcal{P}$~\cite{Abeysekara:2017mjj}, is utilized to efficiently select gamma-like events. 
The $\mathcal{P}$ values are required to be less than 1.02, 0.90, 0.88, 0.88, 0.84, 0.84, and 0.84 for seven $N_{\rm{hit}}$ groups, respectively. These criteria are optimized based on the observed significance of the Crab Nebula. Depending on the energy of the gamma rays and cosmic rays, these criteria can retain roughly more than 60$\%$ of gamma rays while suppressing over 97$\%$ of the cosmic-ray background. Following this background suppression, approximately 44 million gamma-like events are observed by WCDA each day. 

The PBH burst search is conducted by scanning all gamma-like events across three search timescales ($\Delta t=10$~s, 20~s, and 100~s) for significant excesses over the expected cosmic-ray background. All the surviving gamma-like events in each search time duration are filled into the sky map in celestial coordinates (R.A. and decl.), with the sky map divided into $0^{\circ}.1 \times 0^{\circ}.1$  square pixels, according to their arrival direction. The spatial scan for a sky map is performed by sliding a rectangular grid across the sky map until the entire WCDA field of view is covered. The grid size is optimized to $1^{\circ}.2 \times 1^{\circ}.2$ based on the sensitivity study of PBH bursts using the PBH signal MC sample and a background MC sample generated through Poisson sampling from the estimated background map.  The width of the spatial grid in R.A. is scaled with decl. according to $1^{\circ}.2/\cos(\rm{decl.})$ to account for the smaller sky sphere as the decl. moves away from $0^{\circ}$. We continuously slide the time window and repeat the spatial scan until the entire WCDA dataset has been scanned. The step size for sliding the spatial grid (or temporal window) is set to be smaller than the grid size (or the duration of the time window) to ensure a high probability of capturing the majority of gamma-ray events from an individual PBH burst. The total number of search grids ranges from $10^{13}$ to $10^{14}$, depending on the length of the search timescales.

The cosmic-ray background at each spatial grid is estimated using the ‘direct integral method’~\cite{Fleysher:2003nh}. This method assumes that the detection efficiency varies slowly and the time average within a relatively short time window can properly reflect the background. In this analysis, we use a three-hour integration time to estimate the detector acceptance. To minimize statistical fluctuations in background estimation, we require at least 10 events in the corresponding acceptance histogram position for each search grid. However, this requirement is often not met for large $N_{\rm{hit}}$ groups due to the low instantaneous event rate, and thus, it reduces the significance of the PBH burst signal. Therefore, we combine the hit groups with $N_{hit}>100$ and recategorize the data into three groups with $N_{\rm{hit}}$ in ranges of $[30,60),[60,100),$ and $\ge100$. 

We perform a likelihood-ratio test to assess the significance of a burst event across three $N_{\rm{hit}}$ groups within the same grid location. The total test statistics (TS) for each search grid location is calculated as follows:
\begin{equation}
    -2\ln\lambda=\sum_{i=0}^2{2[\ln\mathcal{L}(N_{\rm{obser}})-\ln\mathcal{L}(\rm{B})]},
\end{equation}
where $\mathcal{L}(N_{\rm{obser}})$ or $\mathcal{L}(\rm{B})$ represents the Poisson probability for each grid,  with the mean defined by either the number of the observed events or the estimated background counts. The index $i$ refers the $N_{\rm{hit}}$ groups. Based on Wilks’s theorem, the TS follows a $\chi^2$ distribution with one degree of freedom, and the significance can be calculated as $S=\sqrt{-2\ln\lambda}$.  The $p$-value for each grid is then determined from the computed significance.

\begin{figure}[htpb]
 	\subfigure{
  \includegraphics[trim=0 69 0 0,clip,width=0.99\linewidth]{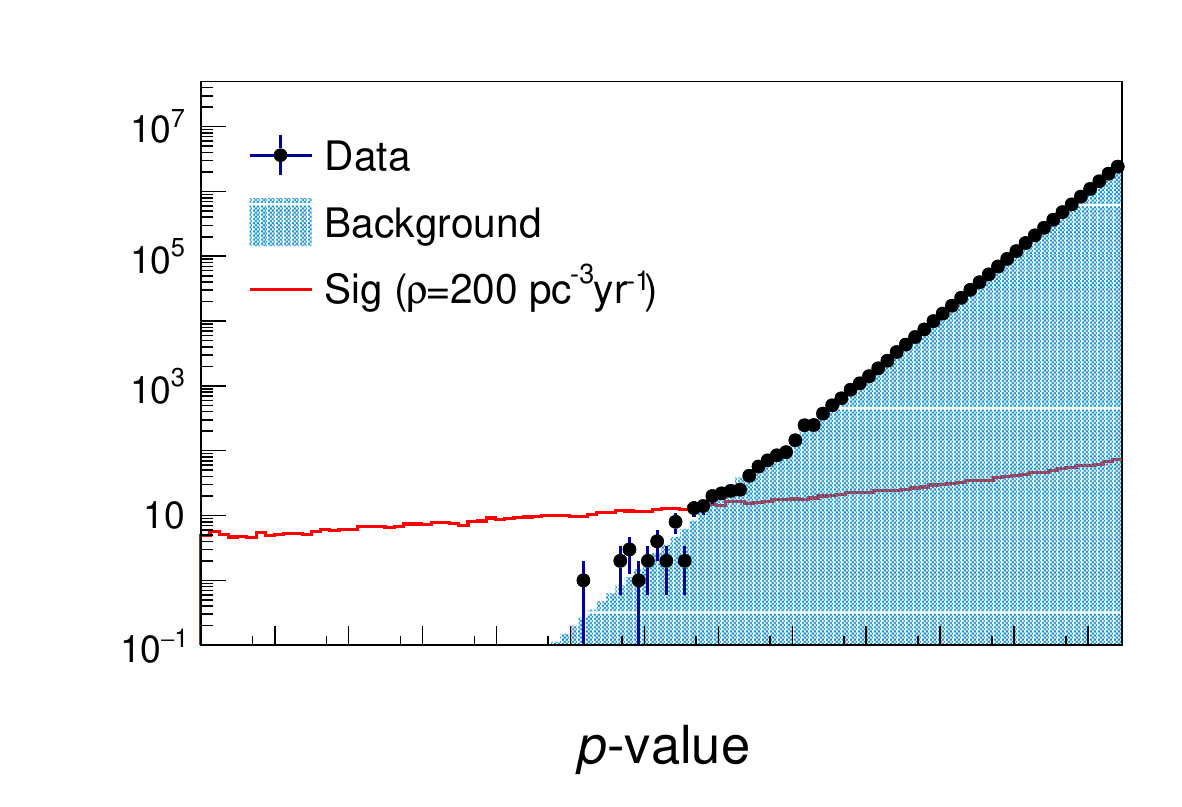}
  \begin{picture}(0,0)
    \put(-110,100){\color{blue}{$\Delta t=10$~s}}
  \end{picture}
  }
  \vskip -6pt
  \subfigure{
  \includegraphics[trim=0 69 0 35,clip,width=0.99\linewidth]{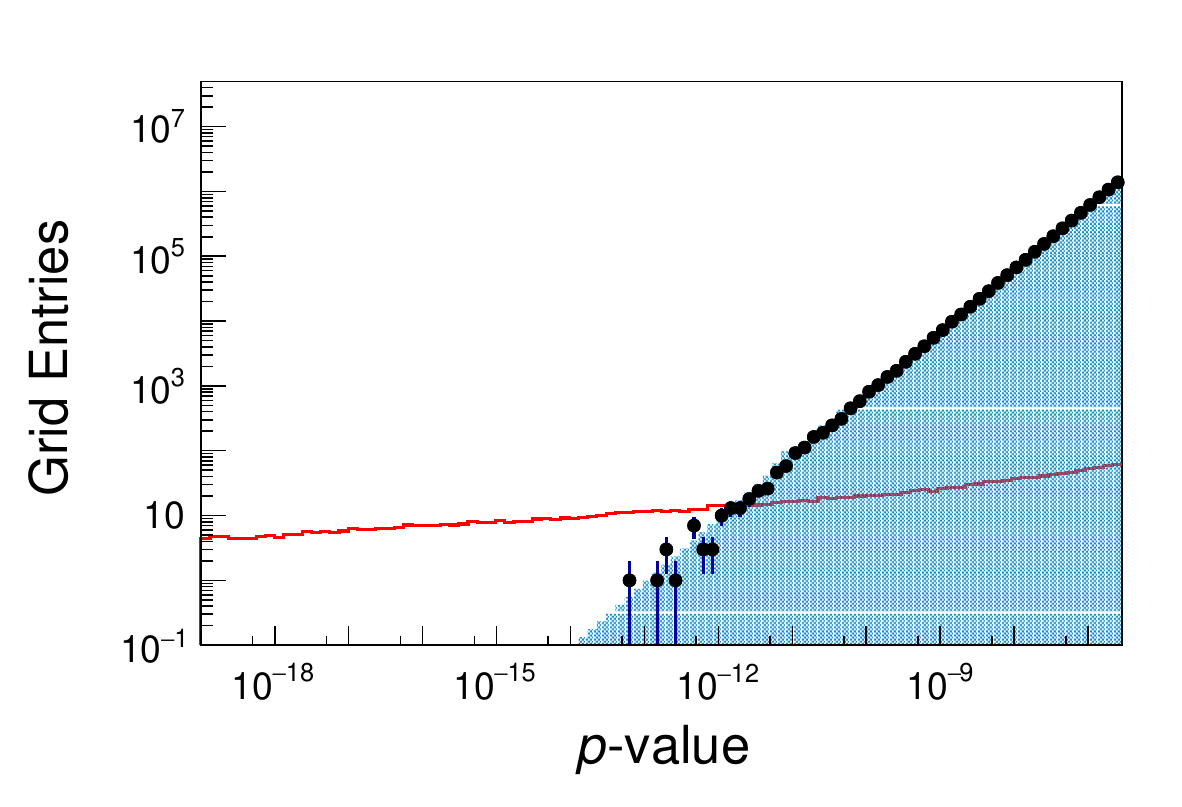}
  \begin{picture}(0,0)
    \put(-180,95){\color{blue}{$\Delta t=20$~s}}
  \end{picture}   
  }
    \vskip -6pt
  \subfigure{
    \includegraphics[trim=0 0 0 33,clip,width=0.99\linewidth]{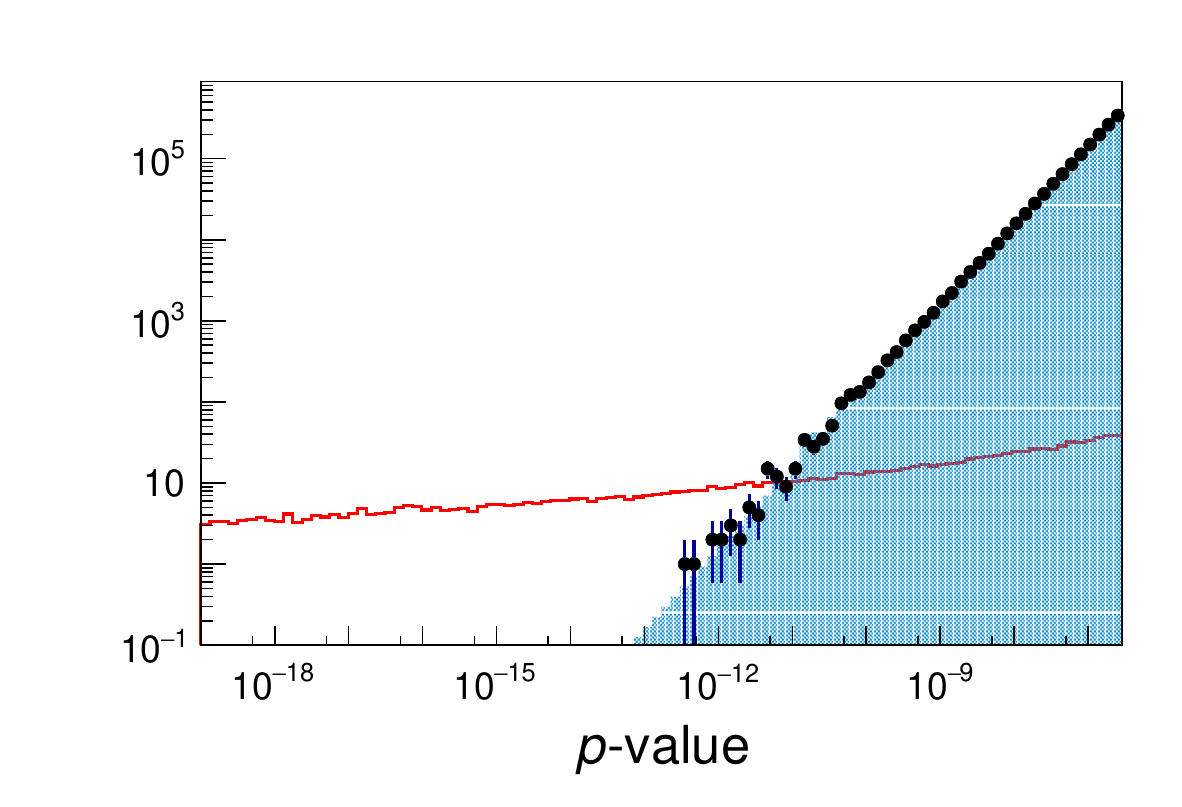}
  \begin{picture}(0,0)
    \put(-180,120){\color{blue}{$\Delta t=100$~s}}
  \end{picture} 
	}
    \caption{The $p$-value distributions for the grids with significance larger than $5\sigma$ across three search timescales of $\Delta t=10$~s, 20~s, and 100~s. The dots with error bars are from the data scan, and the filled blue histograms are from the simulated cosmic-ray background sample. The red histograms are from grids containing gamma-ray events associated with PBH burst signals, derived from a merged sample of both the background simulation and PBH burst signal samples. The local burst rate density of PBH bursts is scaled to $200$~pc$^{-3}$~yr$^{-1}$.}
	\label{fig:pvalue-dis}	
\end{figure}
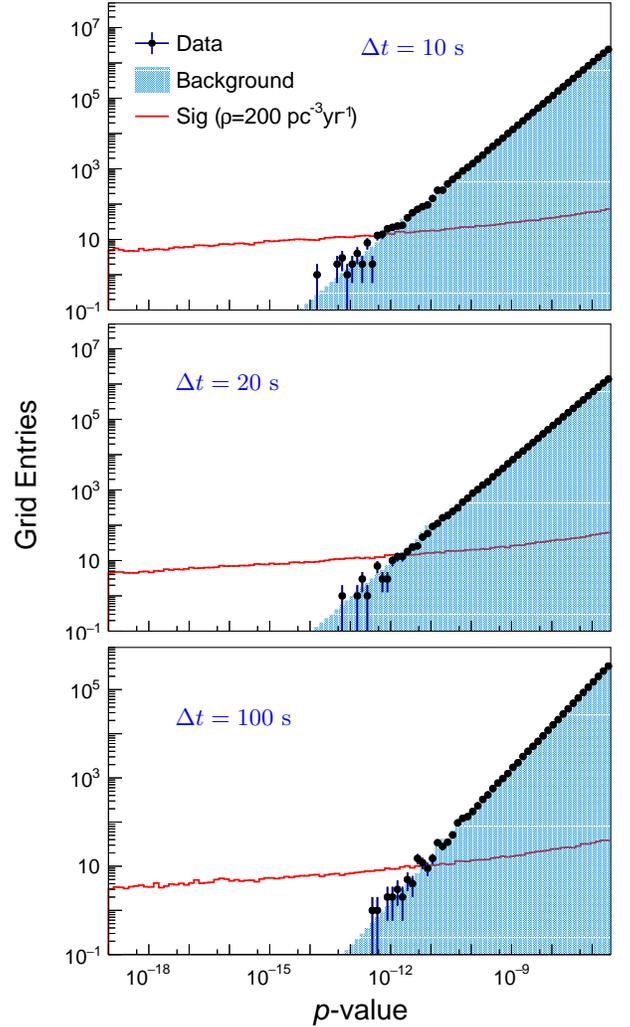

All grids in the data that scan with a significance greater than $5\sigma$ (corresponding to the $p$-value $< 2.87 \times 10^{-7}$) are recorded for further identification of the PBH burst signals. The $p$-value distribution for the recorded grids is expected to consist of two components: cosmic-ray background fluctuations and potential PBH bursts in the data.
To estimate the $p$-value distribution for the cosmic-ray background, we also perform the same scan for the background MC sample. The background sky map for each time window is generated using Poisson sampling based on the estimated background. Since the overlap introduces dependencies between nearby grids and time windows, the grid count cannot directly reflect the number of potential PBH bursts in the data. Therefore, we determine the local burst rate by comparing the $p$-value distribution of all search grids between the data and the MC samples. To determine the $p$-value distributions of PBH burst signals under varying local burst rate density assumptions, we merge the simulated signal MC sample into the background MC samples. The same scanning procedure is applied to the merged samples, and grids containing gamma-ray events from PBH bursts are recorded. The $p$-value distribution of the recorded grids with signal events is considered the $p$-value distribution for PBH signals, as shown by the red line in Fig.~\ref{fig:pvalue-dis}. The existence of PBH bursts is expected to distort the $p$-value distribution across all grids, differentiating it from the background-only distribution, thereby enabling the search for the PBH burst signal using this distinction. 

GRB 221009A is an extraordinarily bright burst and has been observed by LHAASO~\cite{LHAASO:2023kyg,LHAASA:2023pay}. In this work, we also detect GRB 221009A significantly using the blind scan method (details can be found in Supplemental Material~\cite{supp-met}). 
To eliminate its impact on our PBH bursts signal search, we veto the time windows in which GRB 221009A occurs. The resulting $p$-value distributions for recorded grids across three search timescales are shown in Fig.~\ref{fig:pvalue-dis}. 
The signal shapes exhibit a much flatter profile and a higher proportion in the low $p$-value region, making them distinguishable from the background in this range. 
The $p$-value distributions for three search timescales are in good agreement with the expected background shapes, indicating no significant PBH burst detection in the data.

As no significant PBH burst signal is observed in the $p$-value distribution of the data, the upper limit on the local PBH burst rate density is determined at the 99$\%$ confidence level (CL). In our search, the nearby grids have approximately 80$\%$ overlap in both spatial and temporal searches. This overlap causes the grids to be nonindependent, altering the statistical behavior of each $p$-value bin in Fig.~\ref{fig:pvalue-dis}. 
The Poisson distribution expresses the probability of a given number of events occurring in a fixed interval of time. 
We verify the probability density function (PDF) of the background in each $p$-value bin by counting the number of grids within specific $p$-value regions per day from the simulated background sample. The background PDF closely matches the Poisson distribution for the simulated background. 
However, the situation differs significantly when considering PBH signals. 
The PDF for PBH burst signals is determined by simulating the signals thousands of times under the same assumption for the local PBH burst rate density, using the same yearly timescale as the WCDA data. 
The overlap between grids and time windows around PBH burst events introduces strong correlations in the $p$-value distribution, and a long tail appears in the simulated signal PDF, leading to a significant deviation from the Poisson distribution. 
As a result, extracting the upper limit on the local burst rate density based on the Poisson probability assumption for each $p$-value bin is inaccurate. The PDFs for different timescales and $p$-value regions vary significantly, making it difficult to provide a unified analytical description. Therefore, the signal PDFs for different $p$-value regions are derived by simulation.

Furthermore, a binned likelihood fit requires the independence of the $p$-value bins in Fig.~\ref{fig:pvalue-dis}. However, since nearby $p$-value bins are dependent in the case of PBH signals, a binned-likelihood fit to the data is not applicable.
Hence, we select three low-background-level $p$-value regions as a single $p$-value bin for the three search durations to determine the upper limits of the local PBH burst density. The chosen $p$-value regions are $(0-1\times10^{-13})$ for $\Delta t=10$~s, $(0-5\times10^{-13})$ for $\Delta t=20$~s, and $(0-1\times10^{-12})$ for $\Delta t=100$~s, respectively. These ranges are optimized for the best sensitivity of PBH signals based on expected background counts and the simulated signal PDF. The details are described in Supplemental Material~\cite{supp-met}.

\begin{figure}[h]
    \centering
    \includegraphics[width=3.5in,height=3.0in]{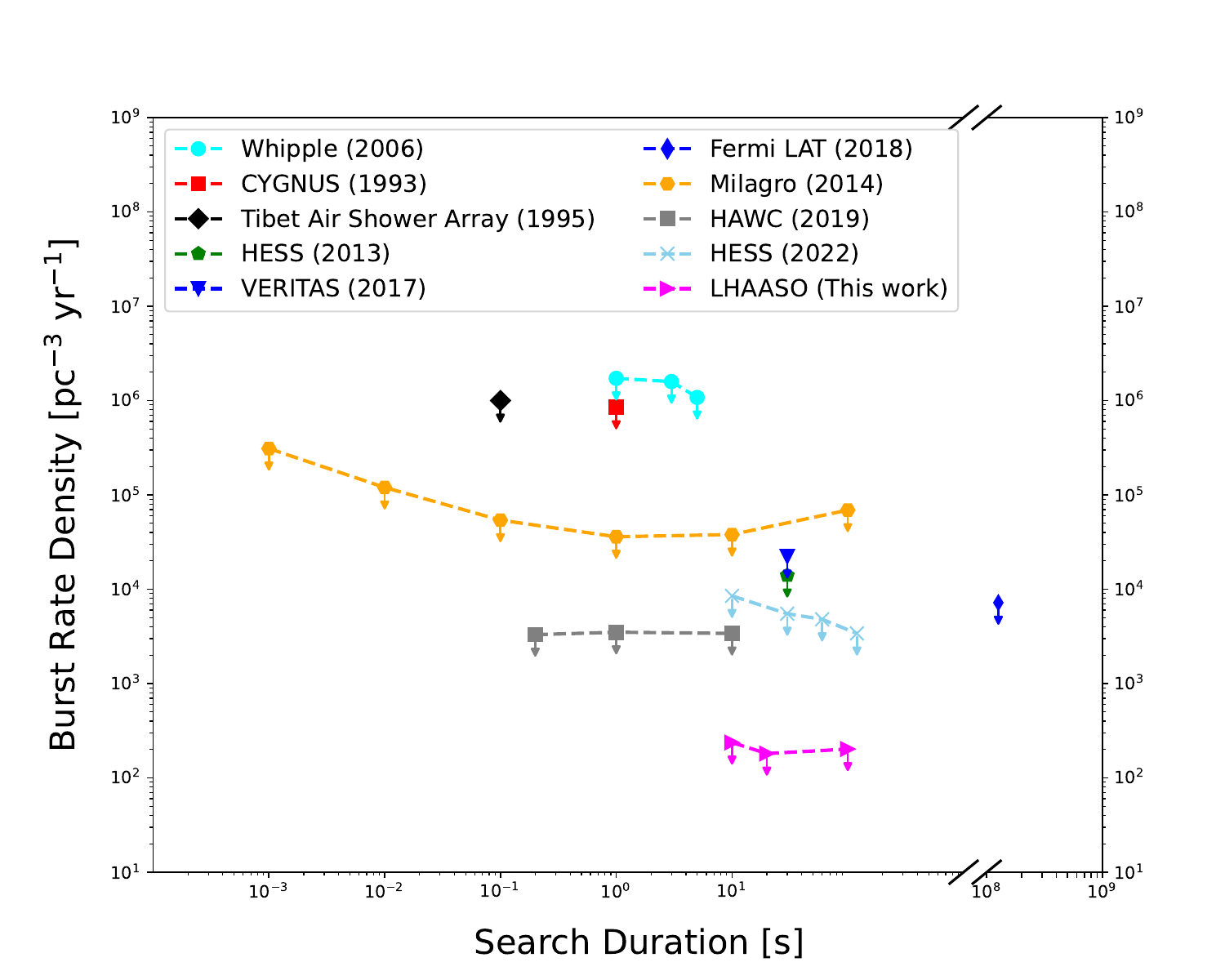}
    \caption{The 99$\%$ CL upper limits on the PBH burst rate density from various experiments. LHAASO provides stringent upper limits on the probe burst timescales of 10 s, 20 s, and 100 s. The previous upper limits from  Whipple~\cite{Linton:2006yu}, CYGNUS~\cite{Alexandreas:1993zx}, HESS~\cite{Glicenstein:2013vha}, VERITAS~\cite{Archambault:2017asc}, \textit{Fermi}-LAT~\cite{Fermi-LAT:2018pfs}, Milagro~\cite{Abdo:2014apa}, HAWC~\cite{HAWC:2019wla}, and HESS~\cite{HESS:2023zzd} are shown for comparison.}
    \label{fig:b_r}
\end{figure}

The probability of the observed grid counts in the three chosen $p$-value regions is calculated by summing over the probabilities of all possible signal and background count combinations. Therefore, the likelihood with a local burst rate density $\rho$ is given by
\begin{equation}
\mathcal{L}(\rho) =\sum_{n_{\text{sig}}=0}^{N_{\rm{obs}}} P_{\text{sig}}(n_{\text{sig}},\rho)\cdot P_{\text{bkg}}(N_{\rm{obs}}-n_{\text{sig}}),
\end{equation}
where $N_{\rm{obs}}$ is the observed count of grids in the given $p$-value regions for the data.  $P_{\text{sig}}(n_{\text{sig}},\rho)$  represents the simulated signal probability for $n_{\text{sig}}$  signal grid counts under the local burst rate density of $\rho$, and  $P_{\text{bkg}}(N_{\rm{obs}} - n_{\text{sig}})$  denotes the probability for the remaining background counts.
To determine the 99$\%$ CL upper limits of the local burst rate density, we perform a likelihood scan under various hypotheses for the local burst rate density. After incorporating systematic uncertainty, the 99$\%$ CL upper limits on the local PBH burst rate density are listed in Table~\ref{tab:up-sys} and shown in Fig.~\ref{fig:b_r}. Compared to previous constraints from other telescopes and observatories, LHAASO demonstrates the highest sensitivity to timescales ranging from several seconds to several hundred seconds.

 \begin{table}[htb]
	\centering
	\caption{The search burst duration and the 99$\%$ CL upper limits on local PBH burst rate density.}
	\label{tab:up-sys}
   \tabcolsep=0.15cm
   \renewcommand{\arraystretch}{1.3}
	\begin{tabular}{c  |   c }
		\hline\hline
		\centering{Search Duration} & 99$\%$ CL Upper Limit (pc$^{-3}$~yr$^{-1}$) \\ 
        \hline
         10 s & 237 \\ 
        20 s & 181\\
         100 s & 202 \\ 
        \hline\hline
	\end{tabular}
\end{table}

$Systematic$ $uncertainty-$The systematic uncertainty in the upper limit measurement primarily arises from the background estimation method and the simulation of PBH burst signals. 
For background estimation, the detector acceptance is estimated using a three-hour integration time, and the background in each grid is then calculated.
To assess the uncertainty in the background histograms, the integration time is varied from three to five hours. Furthermore, we use the $p$-value range of $(10^{-9}-10^{-7})$ as the fiducial region to verify the consistency between the data and the background simulation. The systematic uncertainty from the background estimation is estimated to be 5$\%$ across the three burst timescales.
In the PBH burst simulation, the atmospheric model in the MC simulation, along with variations in atmospheric conditions during operation, affects the detection efficiency, which is not properly modeled in the simulation. This uncertainty is estimated to be as large as 8$\%$ on the effective area~\cite{LHAASO:2023rpg}. Additionally, uncertainties in the PSF influence the number of signals within the given grids, thus impacting the $p$-value distribution of PBH signals. The uncertainty in the simulated PSF is estimated to be $0^{\circ}.05$ by comparing the observed event profiles among Crab, Mrk 421, and Mrk 501~\cite{LHAASO:2023rpg}. To estimate the systematic uncertainty due to the effective area and PSF simulation, we resimulate the PBH burst signals by scaling the effective area by 92$\%$ and enlarging the PSF by $0^{\circ}.05$. The total systematic uncertainty is accounted for by varying the background level within its corresponding uncertainty and using the resimulated signal PDFs to repeat the upper limit measurement. The larger upper limit, compared to the nominal one, is adopted as the final result.

$Conclusion$ $and$ $outlook$$-$In summary, we perform an all-sky search for individual local PBH burst events with burst timescales of 10~s, 20~s, and 100~s, out to a distance of 1~pc using the LHAASO-WCDA data. The PBH emission rate increases significantly in the final few seconds of the PBH lifetime and can be detected through the VHE gamma-ray burst. We search for PBH burst events through the $p$-value distribution across all search grids. No evidence of a PBH burst is observed, and the 99$\%$ CL upper limit on the local PBH burst rate density is determined to be 181~pc$^{-3}$~yr$^{-1}$ with a burst timescale of 20~s. LHAASO provides stringent upper limits on the probed burst timescales, improving the previous best constraint by more than one order of magnitude. 
The construction of the Large Array of Imaging Atmospheric Cherenkov Telescopes (LACT) within LHAASO will significantly improve gamma-proton discrimination and angular resolution~\cite{Zhang:2023bsi}. With LACT and increasing data, LHAASO is expected to deliver more competitive results on PBH burst events, advancing the study of extreme astrophysical phenomena.

$Acknowledgments$ $-$We would like to thank all staff members who work at the LHAASO site above 4400 meters above sea level year round to maintain the detector and keep the water recycling system, electricity power supply, and other components of the experiment operating smoothly. We are grateful to Chengdu Management Committee of Tianfu New Area for the constant financial support for research with LHAASO data. We appreciate the computing and data service support provided by the National High Energy Physics Data Center for the data analysis in this paper. This research work is supported by the following grants: The National Natural Science Foundation of China  No. 12175248, No. 12447105, No. 12175243, No. 12473001, No. 11975072,
No. 12393851, No. 12393852, No. 12393853, No. 12393854, No. 12173039, No. 12261160362, the National SKA Program of China (grant Nos. 2022SKA0110200, 2022SKA0110203), the National Key R\&D Program of China No. 2023YFC2206403, the Department of Science and Technology of Sichuan Province, China No.24NSFSC2319, No.2024NSFJQ0060, Project for Young Scientists in Basic Research of Chinese Academy of Sciences No.YSBR-061, and in Thailand by the National Science and Technology Development Agency (NSTDA) and the National Research Council of Thailand (NRCT) under the High-Potential Research Team Grant Program (N42A650868).

\end{document}